# Review of surface discharge experiments.

**Bloshchitsyn Vladimir**

**St.-Petersburg State University, Physics Faculty**

## Introduction

Experiments on the surface discharge were described in the literature ( [1], [2]). The surface discharge can slide on low-conductivity material (PTFE, PMMA, ceramics, polycarbonate) and conductors ( water, ice, earth, ... ). These and intermediate variants can be presented with simple scheme (see Fig. 1). There is surface discharge with capacitive energy source, when $C_1$ large, if $R_1$ small , then surface discharge developed due by surface conduction. Capacitance $C_1$, depend on dielectric permittivity of material $\epsilon$, geometry of dielectric (position, thickness) and electric field configuration. Resistance $R_1$ depend on surface conductance which associated with conductivity σ of material and with state of surface.

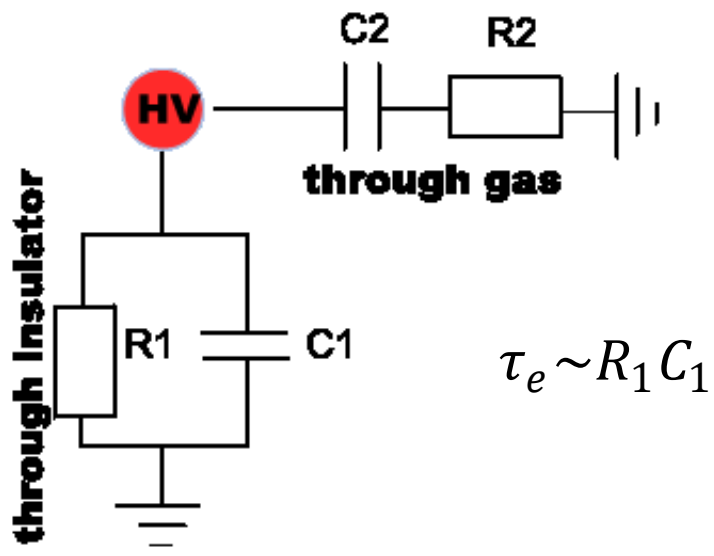

Fig. 1 Conditional schema that describes and sliding discharge, and the discharge of conducting surfaces

| | | |
|---|---|---|
| SURFACE DISCHARGE | 1. MEDIUM | 1.1 VACUUM |
| | | 1.2 LIQUID DIELECTRICS |
| | | 1.3 GASES |
| | 2. DIELECTRIC PROPERTIES | 2.1 CONDUCTIVITY $\sigma(\omega)$ |
| | | 2.2 PERMITIVITY $\epsilon(\omega)$ |
| | | 2.3 WETTABILITY $\alpha$ |
| | | 2.4 SMOOTHNESS |
| | 3. COMPUTER MODELS | 3.1 CORONA |
| | | 3.2 STREAMER |
| | | 3.3 SPARK, ARC |
| | 4. EXPERIMENTS | 4.1 ELECTRIC LINES PARALLEL TO SURFACE |
| | | 4.2 ELECTRIC LINES PERPENDICULAR TO SURFACE |
| | | 4.3 COMPLEX CONFIGURATIONS |
| | 5. APLICATIONS | 5.1 LAMPS |
| | | 5.2 ARRESTERS, SWITCHES |
| | | 5.3 BUSHINGS, POST INSULATORS |
| | | 5.4 WATER, AIR TREATMENT |
| | | 5.5 FLOW CONTROL |
| | 6. EFFECTS | 6.1 TRIPPLE JUNCTION |
| | | 6.2 COLLECTIVE EFFECT |
| | 7. TYPE OF INTERACTIONS | 7.1 SURFACE CONDUCTIVITY |
| | | 7.2 POLARIZATION |
| | | 7.3 CHARGE ACCUMULATION |
| | | 7.4 PHOTOIONIZATION |
| | | 7.5 ION, ELECTRON EMISSION |
| | | 7.6 DESORPTION |
| | 8. TYPE OF APPLIED VOLTAGE | 8.1 PULSED |
| | | 8.2 AC |
| | | 8.3 CONSTANT |
| | 9. VALUE OF CURRENT | 9.1 LOW (CORONA, STREAMER) |
| | | 9.2 HIGH (SPARK, ARC) |

Tab. 1 Different aspects of surface discharge phenomena

Different aspects of surface discharge is enumerated in Table 1.

## Surface discharge applications

Surface discharge is big problem for high-voltage equipment (rigid insulators and bushings of outdoor switchgear covered by moisture are subjected to breakdown). There are several terms in literature for surface discharge. Sliding discharge and creeping discharge are equivalent terms. Surface discharge is named dielectric barrier discharge, when one or both electrodes are covered by insulator. DBD is base for Plasma Displays Panel (PDP) (p. 682 [3]). Practical interest in the surface discharge is manifested not only as a negative phenomenon in high-voltage technology ( [4], [5] ), also surface discharge can be used to construct an effective earthing ([2]). Surface discharge is used for water purification with streamer corona technology [6]. It is also proposed to use the barrier discharge with an isolated electrode for flow control in aerodynamic applications ([7]). The breakdown voltage in a surface discharge is less then gas (vacuum) discharge ( thanks to a charge carriers formation near the dielectrics (see diagram Fig.8 ), as well as the effect of the so-called triple junction ([8], [9]). It's of interested for High Intensity Discharge [HID] lamps technology, which arc initiation voltage should be minimal [10]. Surface discharge also has been studied in a vacuum, in the context of the establishment of effective metal-cathode used for production of large-cross-section beams (LCSB), which can be used for pumping gas lasers ([8]). Also breakdown along the dielectric surface in vacuum is a limitation in increasing the power of pulsed power devices. Interesting insulators are developed (see Fig. 2) with alternating metal and dielectric inserts to increase the electric strength of vacuum devices ([11], [12]).

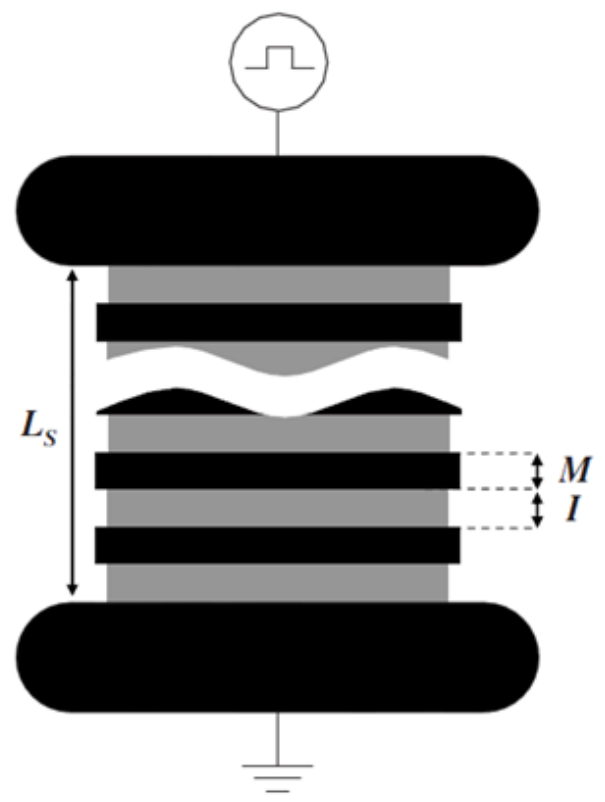

Fig. 2 Multilayer vacuum insulator schema [13]

## Classification of experimental works on surface discharge

Surface discharge is studied in different interelectrode gaps: in **vacuum** ([8],[13], [14]), in **liquid dielectrics** ([15], [16]), in **gases** ([17], [18], [19], [20]), in **artificial** and in **ambient air** ([21]). For mediums in which the development of the discharge in normal conditions is difficult (vacuum and liquid dielectrics), the insulator surface should play a greater role than in gases, because surface can be a source for the discharge plasma. Generality of mechanisms of surface discharges in different mediums is in interaction of sparks and streamers with the surface of dielectrics (see later Fig. 8).

## Experimental setups for surface discharge investigation

Some type of experimental setups is dictated by practical device, which surface discharge was investigated (for example, HID lamps [10] Fig. 3, flow control devices [7], water purification devices [6]).

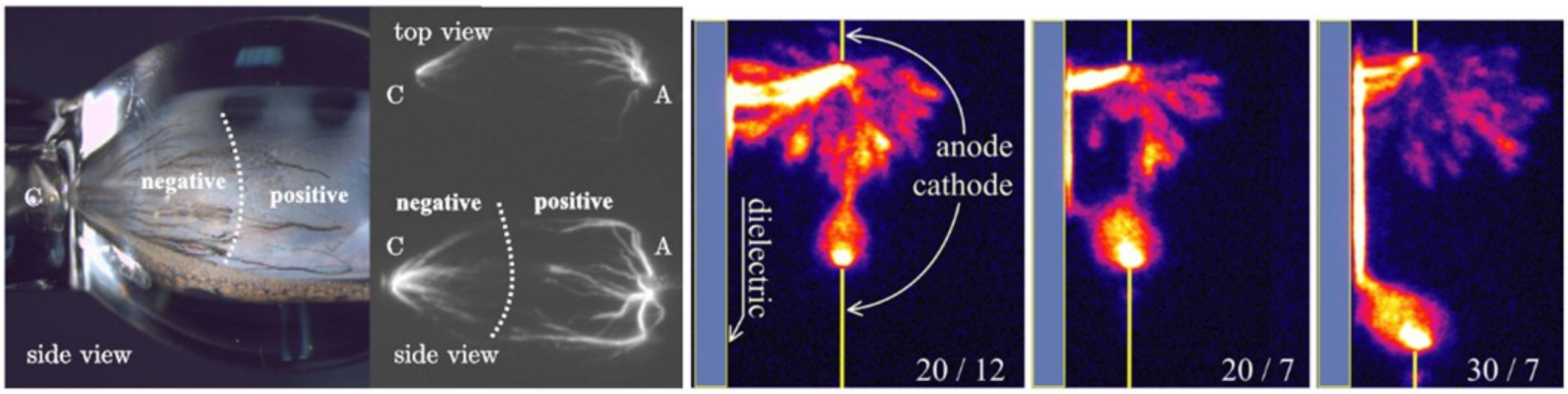

Fig. 3  Discharge in HID lams [12] (left) и [20] (right).

These configurations are complex for comprehension and, therefore, it's needed simplest experimental setups for surface discharge phenomena investigation.  Experimental setups to study the surface discharge can be divided into 2 groups (Fig. 4).

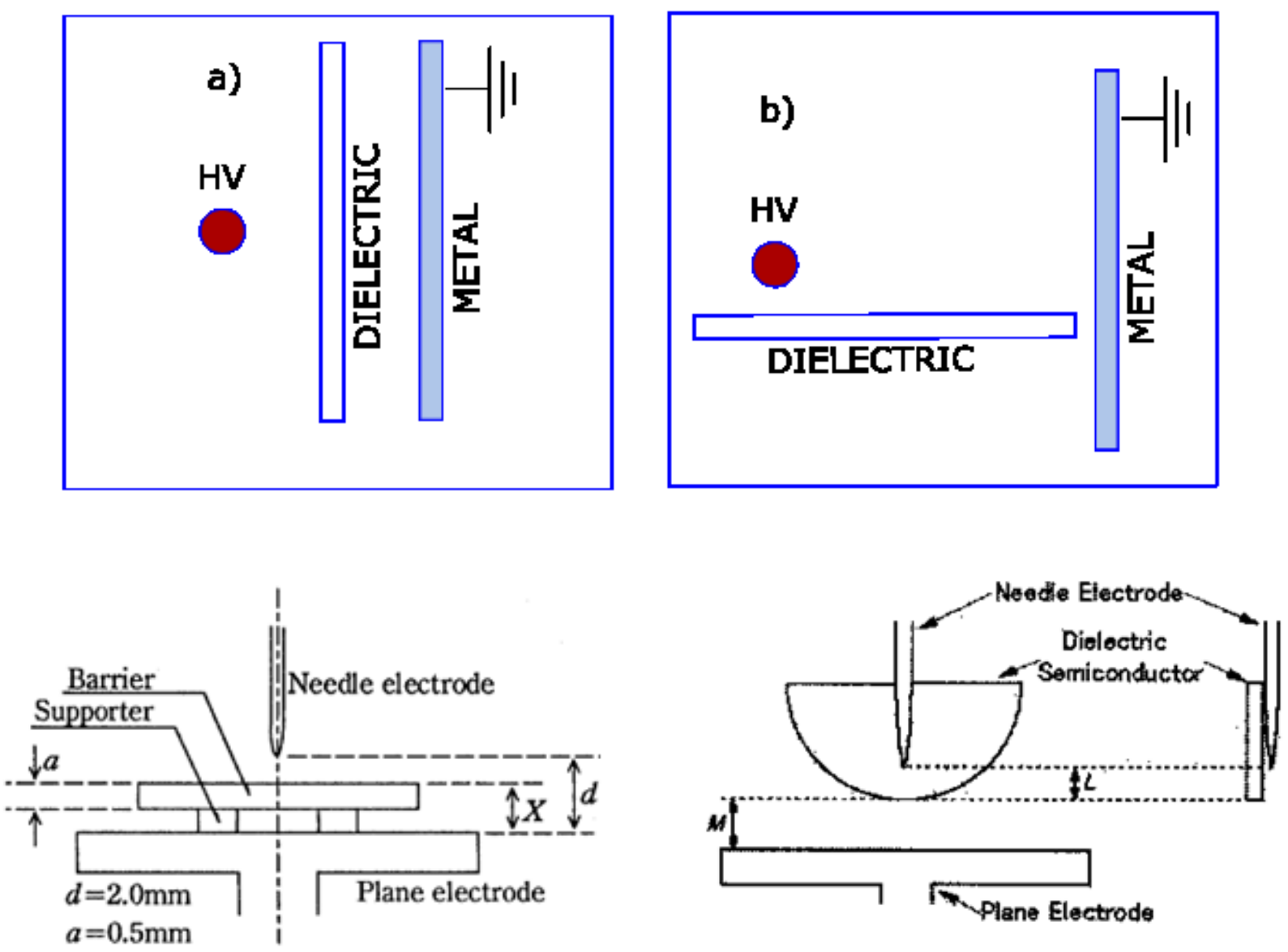

Fig. 4 Common types of experimental setups are used the surface discharge investigation and their specific implementation [19] (left - bottom) [21] (right - bottom)

Setups type a) are used in papers: [17], [22], [23],[24], [18], [15] type b): [5], [4], [19]. Experimental setups type a) and b) can transform into each other with an appropriate choice of the experimental setup configuration, for example Fig. 5 when the angle α from 45 to 90 degrees the scheme type a) will transform into the scheme type b), and the ratio $C_1/R_1$ (in common schema Fig. 1) will vary.

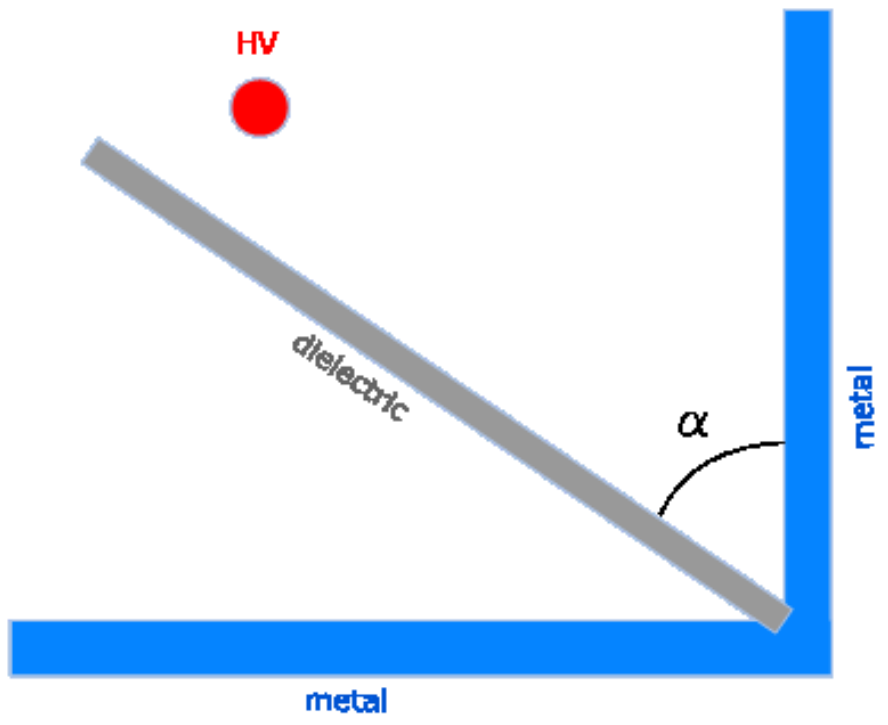

Fig. 5 Thought experiment for surface discharge investigation is shown (when α → 90 capacitive mechanism of the discharge will be significant).

Truncated version of this method was used in [21], which slope of the dielectric surface adjacent to the electrode needle was varied.

## Data is obtained in the experiments

Data are obtained from surface discharge experiments are consist of: 1) video frames (including high-speed photography) 2) the pulse shape of the applied voltage 3) oscillograms of electric current 4) the surface charge density and electric field on the surface of the insulator 5) dust (Lichtenberg) figures. Unfortunately, there is no works on the spectroscopic investigation of surface discharge plasma in the reviewed literature, although these data could clarify the role of insulators in the formation of the surface discharge.

Often the authors have not paid adequate attention to the surface charge [21] although, in my view it was he who determines the behavior of streamers near the insulators. In [17] [19] the authors as the timing between experiments is used the Maxwellian relaxation time $\epsilon_0 \epsilon \rho$ for surface charge neutralization, but even if the surface charge is absent before experiments, during the voltage pulse it is the accumulated on surface (similar to Fig.6). It's changed the electrostatic configuration of the experiment, so in my opinion, in many situations, the surface charge has a key role.

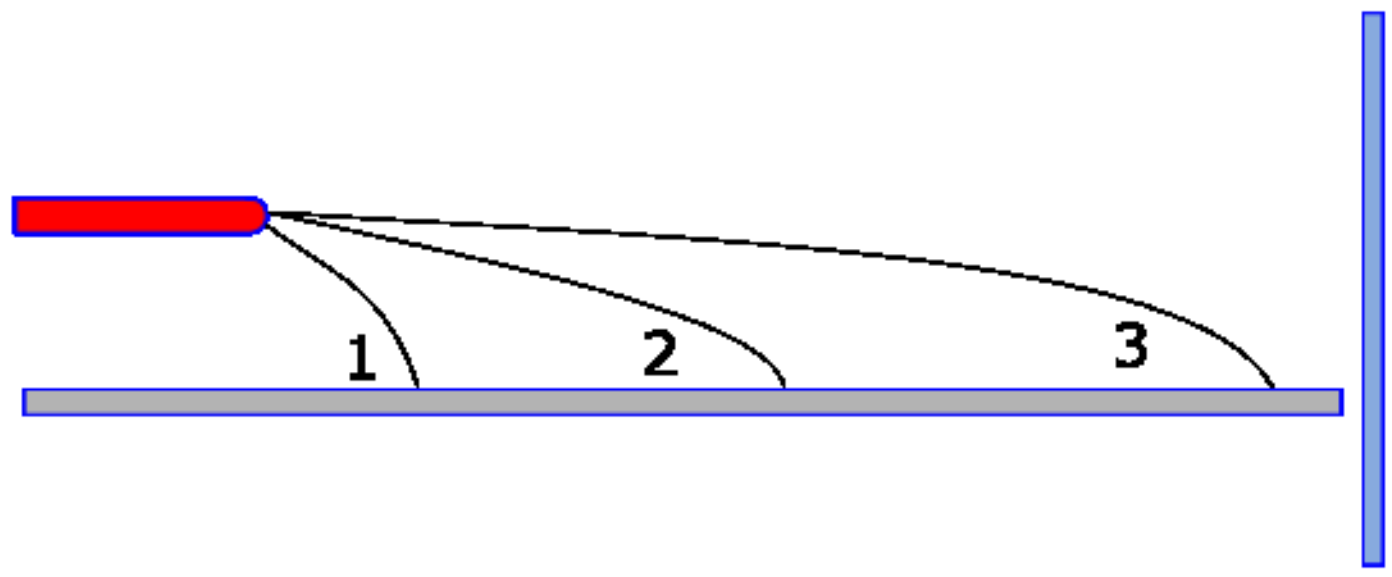

Fig. 6 Streamer "touches" dielectric surface and leaves trace of charge (progress with time - 1), 2), 3))

Often, the authors measure the speed of discharge ([20]), or area of distribution ([21], [19]). The authors proceed from the hypothesis that near the surface discharge develops faster ([20]), due to the processes of plasma formation from dielectric material. Area of distribution is measured by video frames (with iCCD camera, rarely CCD), while the question of error in determining the distance requires explanation. Other methods is based on measuring the electric field near surface ([22], [24]) , or determining size of the Lichtenberg figures. Capacitive probes with precision positioning[24], a variety of fluxmeters (Field Mills), Pockels sensors [22] ( Fig. 7) are used for surface charge density measurements. Lichtenberg figures provide the most visual data, but there are difficulties in interpretation and quantitative analysis of pictures. Other methods offer no such visible results, but they allow you to receive value of the surface charge densities.

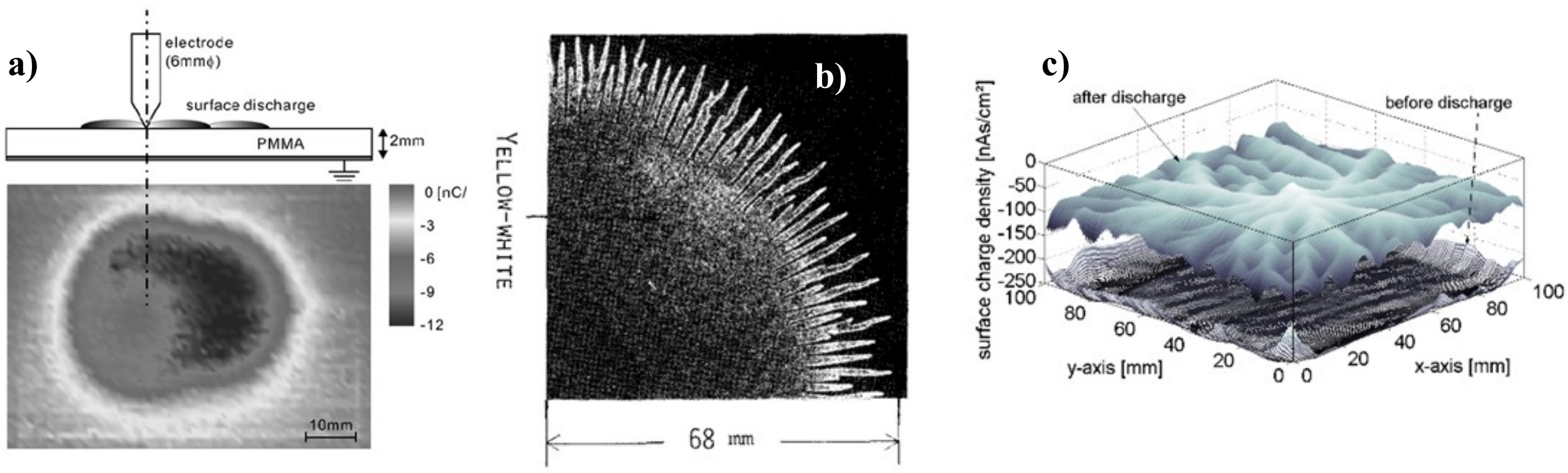

Fig. 7 There is different methods for surface charge measurements and visualization: Pockels sensor a)[24], b) Lichtenberg figures [20], c) Capacitive probe [26].

### Interaction of streamers and sparks with the surface

Possible mechanisms of streamer and spark interaction with the surface illustrated in Fig.8. Some of the mechanisms plays the greatest role in each particular mediums and material of dielectric . Task of the experimenter to determine which the mechanism of interaction with the surface plays a crucial role.

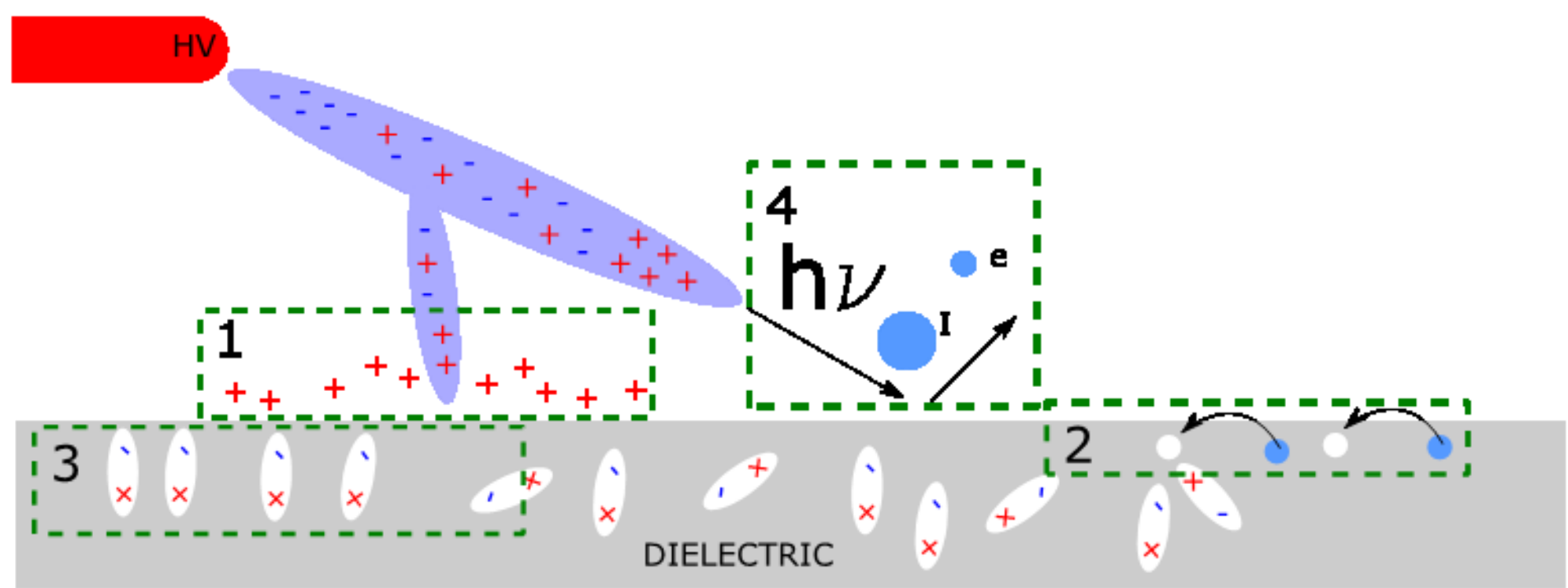

Fig. 8 There are interaction streamer and spark with the surface. Possible mechanisms are: 1 - through surface charge, 2 - through the surface conductivity, 3 - through the polarization of the insulator, 4 - through photoemission and thermionic emission

Interaction of streamers and sparks through the photoemission and the charge on the surface is described in the literature (relevant references are given in [20]). In our view the interaction of streamers and sparks with the surface is of a different nature (see Fig. 9).

Streamer head tracks retrace electric field lines type 1 and 2 (Fig. 9) is observed in experiments [21].

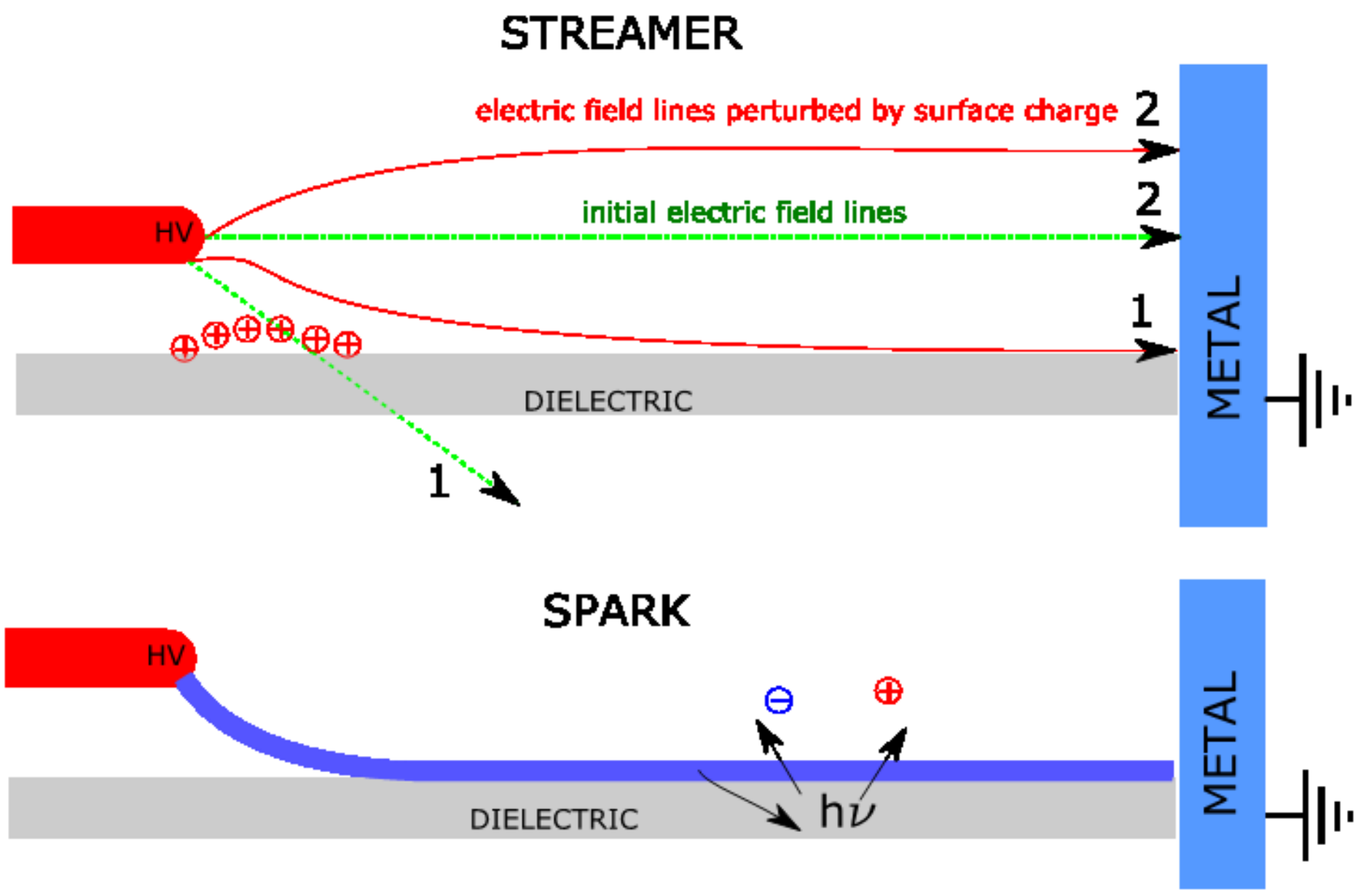

Fig. 9 There is the difference in the interaction between the streamer and spark with the surface

## Influence of dielectric material

In the experiments described in the literature, in addition to geometric parameters of dielectrics (thickness, length), conductivity and electric permittivity values of materials are varied [19]. Some authors select insulators by varying ε (see Fig 10) ([20], [21]), others change ρ (see Fig.11) [5], [4]. The authors [20] concluded that varying of electric permittivity value has weak influence (they used optical glass BK7 with electric permittivity 2.3-4.5 and aluminum oxide with ε = 9.1).

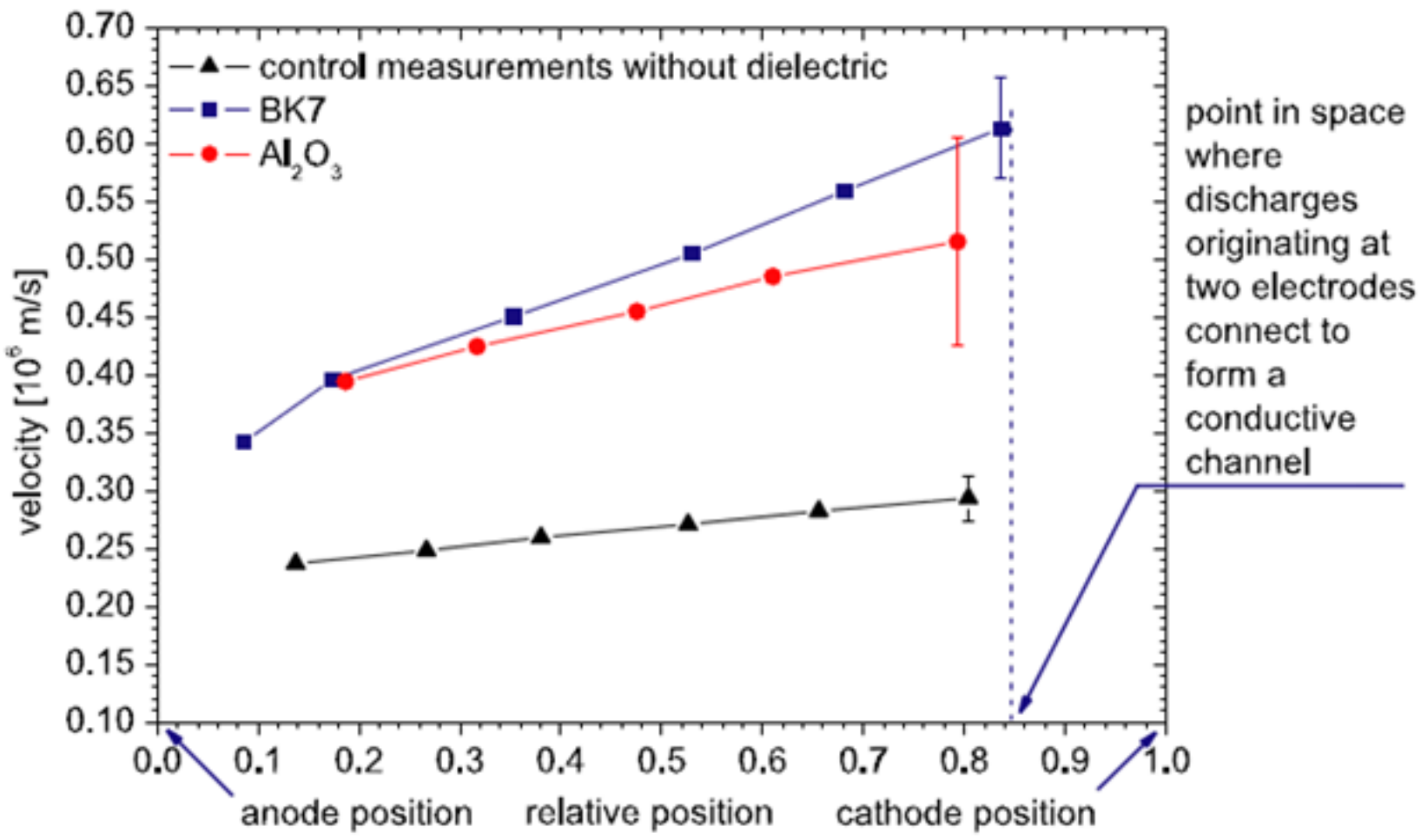

Fig. 10 On the influence of electric permittivity constant on the surface discharge development [22]

This conclusion stems from the fact that the varying of electric permittivity value couldn't significantly increase the capacity $C_1$ in schema Fig. 1. Effect would be greater, in the event of experimental configurations such as Fig 4 a). In the case of variation of the conductivity Fig. 11, the authors showed the influence of the conductivity, although the data are without error estimation, and for more conductive materials trend is unclear.

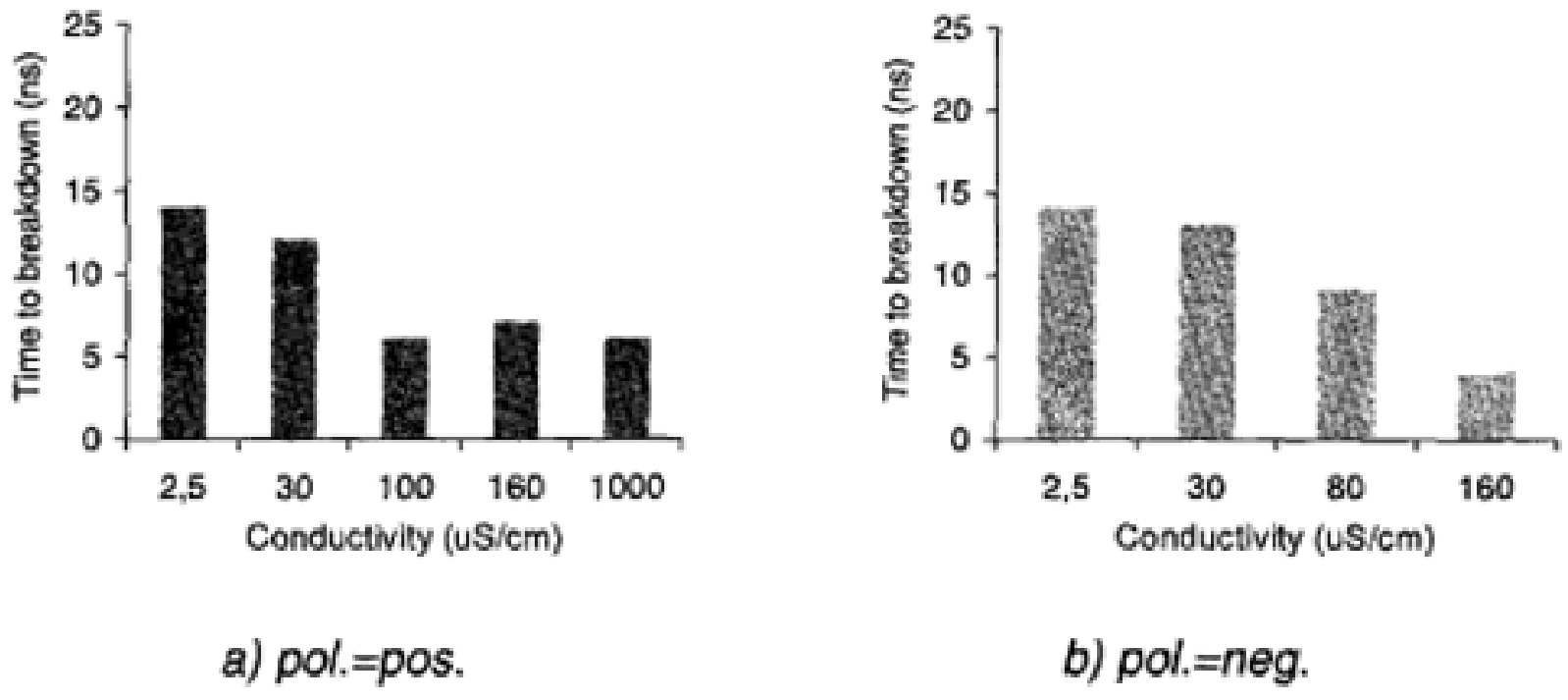

Fig. 11 On the influence of conductivity on the surface discharge development [5]

It is possible that the impact on the surface discharge except for ε and ρ affects the specific surface conductivity, which, for example, in experiments in ambient air can depend on humidity and atmospheric properties.

### Conclusion

As can be seen from the review, the experiments are divided into 2 groups, as if independent experiments, and thus relation lost between interconnected experiments (like Fig.5). Not always attention is paid to the surface charge in the experimental works. In some experimental setups capacity $C_1$ (Fig. 1) is small and the variation of ε cannot significantly change their value. Spectroscopic investigation of surface discharge plasma needed for clarifying role of dielectrics.